\documentclass[a4paper,11pt]{article}
\usepackage{pos}
\usepackage{multicol}
\usepackage{multirow}
\usepackage{colortbl}
\usepackage{tikz}
\usepackage{tikz-feynman}
\usepackage{booktabs} 
\newcommand{\halfcolor}[3]{%
  \tikz[baseline=(char.base)]{
    \node[inner sep=0pt] (char) {\textcolor{#2}{#3}};
    \clip (char.south west) rectangle ([yshift=-0.7ex]char.north east);
    \node[inner sep=0pt, text=#1] at (char.center) {#3};
  }%
}
\tikzstyle{quark}=[thick,postaction={decorate,decoration={markings,mark=at position 0.5 with {\arrow{>}}}}]
\tikzstyle{photon}=[thick, decorate, decoration={snake, amplitude=0.5mm, segment length=1.5mm}]
\tikzstyle{vertex}=[thick, black,fill=white]

\tikzstyle{quark}=[postaction={decorate,decoration={markings,mark=at position 0.5 with {\arrow{>}}}}]
\tikzstyle{photon}=[decorate, decoration={snake, amplitude=0.2mm, segment length=.7mm}]
\tikzstyle{vertex}=[draw=black,fill=white]
\tikzstyle{externalVertex}=[draw=black,fill=black]
\newcommand{\drawTriangle}[2]{%
    \pgfmathsetmacro{\h}{0.3}
    \draw[fill=red, draw=black]
    ($(#1)+(-0.5*\h,-0.289*\h)$) --
    ($(#1)+(0.5*\h,-0.289*\h)$) --
    ($(#1)+(0,0.577*\h)$) -- cycle;
}
\newcommand{\drawPhotonInsertion}[1]{%
    \draw[draw=black, fill=green] ($(#1)-(.1,.1)$) rectangle ($(#1)+(.1,.1)$);
}
\newcommand{\drawDoublePhotonInsertion}[1]{%
    \draw[draw=black, fill=blue!50!cyan, rotate=45] ($(#1)-(.1,.1)$) rectangle ($(#1)+(.1,.1)$);
}
\newcommand{\drawVertexPhotonInsertion}[1]{%
    \node[regular polygon, regular polygon sides=5, draw=black, fill=orange,
        minimum size=4pt, inner sep=0pt, outer sep=0pt] at (#1) {};%
}
\newcommand{\drawDoubleVertexPhotonInsertion}[1]{%
    \node[star, star points=5, draw=black, fill=yellow,
        minimum size=4pt, inner sep=0pt, outer sep=0pt] at (#1) {};%
}

\definecolor{green}{RGB}{79, 141, 16}

\definecolor{mag}{RGB}{141, 16, 79}

\definecolor{myNewColorA}{RGB}{16,79,141}
\definecolor{myNewColorC}{RGB}{21,105,187}

\newcommand{\rcstar}{RC$^{\star}$~}
\newcommand{\cstar}{$C$-periodic~}

\newcommand{\uspin}{$a_\mu^\mathrm{U}$}
\newcommand{\uspinW}{$a_\mu^\mathrm{U,w}$}

\newcommand{\RM}{$\mathrm{RM123}$}
\newcommand{\gcdqedC}{$\mathrm{QCD+QED}$}

\ShortTitle{Comparing RM123 and non-perturbative QCD+QED approaches to the HVP}
\title{Comparing RM123 and non-perturbative QCD+QED approaches to the HVP with C-periodic boundary conditions}

\FullConference{The 42nd International Symposium on Lattice Field Theory (LATTICE2025)\\
2-8 November 2025\\
Tata Institute of Fundamental Research, Mumbai, India\\}

\author[a]{Anian Altherr}
\author[b]{Isabel Campos}
\author[c,d]{Alessandro Cotellucci}
\author[a]{Roman Gruber}
\author[a]{Tim Harris}
\author[a]{Javad Komijani}
\author[e,f]{Francesca Margari}
\author*[a]{Marina K. Marinkovic}
\author[a]{Letizia Parato}
\author[c,g]{Agostino Patella}
\author[b]{Sara Rosso}
\author[e,f]{Nazario Tantalo}
\author[a]{Paola Tavella}

\affiliation[a]{Institut für Theoretische Physik, ETH Zürich, Wolfgang-Pauli-Str. 27, 8093 Zürich, Switzerland}
\affiliation[b]{Instituto de Física de Cantabria and IFCA-CSIC, Avda. de Los Castros s/n, 39005 Santander, Spain}
\affiliation[c]{Humboldt Universität zu Berlin, Institut für Physik and IRIS Adlershof, Zum Großen Windkanal 6, 12489
Berlin, Germany}
\affiliation[d]{Jülich Supercomputing Centre, Forschungszentrum Jülich, D-52428 Jülich, Germany}
\affiliation[e]{Dipartimento di Fisica, Università di Roma “Tor Vergata”, Via della Ricerca Scientifica 1, I-00133 Roma, Italy}
\affiliation[f]{INFN, Sezione di Roma “Tor Vergata”,
Via della Ricerca Scientifica 1, I-00133 Roma, Italy}
\affiliation[g]{DESY, Platanenallee 6, D-15738 Zeuthen, Germany}

\emailAdd{marinama@ethz.ch}
\abstract{Isospin-breaking corrections to the HVP are among the leading sources of uncertainty in the Standard Model prediction of the muon $g-2$. In recent work by the \rcstar collaboration, we compute the intermediate window contribution for a flavour non-singlet current using two strategies to include isospin-breaking corrections: the RM123 approach and a fully non-perturbative dynamical QCD+QED simulation. In both computations, we use $C$-periodic spatial boundary conditions to ensure that locality, gauge invariance, and translational invariance are preserved throughout the calculation. At fixed lattice spacing and volume with $N_f =1+2+1$ dynamical fermions, and fully including sea-quark effects in both computations, we find that simulating the full QCD+QED distribution yields smaller uncertainties for a fixed statistics. We summarize the comparison of the two approaches and discuss the implications for future lattice QCD+QED computations.
}

\begin{document}
\maketitle

\section{Motivation for comparing non-perturbative $\mathrm{QCD+QED}$  and \RM~approach}
Achieving sub-percent precision in lattice computations of hadronic observables requires accounting for isospin-breaking (IB) corrections originating from the difference in masses and charges of the up and down quarks. In particular, the need for improved theory predictions of the anomalous magnetic moment of the muon~\cite{WP25} has made the reliable inclusion of QED and strong IB corrections a central goal of present lattice QCD calculations. This motivates the development of lattice formulations of QCD+QED that maintain the fundamental properties required for controlled continuum extrapolations. A formulation of QCD+QED implemented by the  
\rcstar collaboration is based on \cstar boundary conditions~\cite{Kronfeld:1990qu, Lucini:2015hfa}.   
This choice allows us to perform dynamical simulations of QCD+QED in a finite box while preserving locality, gauge and translational invariance. 
Generation of \cstar non-perturbative QCD+QED configurations is computationally demanding~\cite{openQxDP}, but provides a conceptually clean approach in which QCD and QED effects are treated on the same footing. The \rcstar collaboration has thus far focused on observables where IB corrections are phenomenologically relevant, including charged meson masses, baryons~\cite{openQxDP2, SaraProceedings}, and hadronic contributions related to the anomalous magnetic moment of the muon (muon $g-2$)~\cite{LetiziaProc,PaolaP}. 

A commonly applied strategy for including IB effects is the \RM~method~\cite{RM123}, where IB corrections are computed using conventional dynamical QCD configurations by perturbatively expanding in electromagnetic coupling $\alpha$ and $m_u-m_d$ around isospin-symmetric QCD. This method has proven efficient in many muon $g-2$ related applications~\cite{WP25}, 
and was compared to quenched QED approach 
in the context of hadronic vacuum polarization (HVP) contribution to muong $g-2$ for Domain Wall fermion action~\cite{Boyle:2017gzv}. Here, we report on the comparison of the  non-perturbative QCD+QED approach to the \RM~method, both applied to gauge ensembles with $C$-periodic spatial boundary conditions corresponding to a lattice spacing $a\approx0.05\:\mathrm{fm}$ and the same
statistics, i.e.,  the same number of gauge configurations, $N_{cnfg}=2000$, the same number of point sources, $N_s = 4$, per configuration, and fully including
sea-quark effects in both approaches. 

Our goal was to perform a same statistics comparison between non-perturbative QCD+QED simulations and the {\RM}~approach 
in a controlled setup where the simulation parameters are matched as closely as possible. The observable chosen for this comparison is the intermediate window contribution to the $U$-spin symmetric leading HVP, which we denote as \uspinW. 
The comparison is carried out on two ensembles with 
unphysical quark masses and comparable lattice parameters, differing only in the inclusion of QED and in the $m_u-m_d$ splitting. The first ensemble corresponds to non-perturbative QCD+QED with $m_u \neq m_d$, while the second ensemble is generated in isospin-symmetric QCD and used together with the \RM~expansion. This setup allows for a direct comparison of statistical precision of the two approaches, with an aim to guide future lattice  calculations of precision hadronic observables that require the inclusion of IB effects. Early results of this comparison, obtained without including sea-quark effects, were presented in~\cite{LetiziaProc}. A more detailed account of the updated analysis discussed in these proceedings can be found in~\cite{PaolaP,PaolaThesis}.

\section{Lattice action and choice of observable for comparison}

The gauge ensembles used in this work are generated with $O(a)$-improved Wilson fermions and $N_f=1+2+1$ dynamical quark flavors, where $m_d = m_s$. This choice corresponds to $U$-spin symmetry and allows us to 
choose an observable closely related to the HVP contribution to the muon's $g-2$. This observable is defined from the correlator of the $U$-spin vector current
\begin{align}
V_\mu(x) = \frac{1}{2}\left[\bar{s}(x)\gamma_\mu s(x) - \bar{d}(x)\gamma_\mu d(x)\right],
\label{eq:non-singlet}
\end{align}
which remains flavour non-singlet for $m_d = m_s$ even when isospin symmetry is explicitly broken, as in the setup used in our comparison of non-perturbative QCD+QED and the perturbative \RM~approach. Because of its flavour structure, the two-point function corresponding to the current defined in Eq.~(\ref{eq:non-singlet}) receives only valence quark-line connected contributions, making it significantly simpler to compute than the full electromagnetic current correlator. In the SU(3)-symmetric limit, this correlator is proportional to the light-quark HVP contribution, $a_\mu^{\mathrm{U}}=\frac{3}{4} a_\mu^{\mathrm{uds}}$
~\cite{Gerardin:2018kpy}.

\uspin~is computed in the time-momentum representation~\cite{BerneckerMeyer}, and we focus first on the intermediate time window, dominated in the range $[0.4,1.0] \mathrm{fm}$, where both finite volume and discretization effects are suppressed. The resulting contribution to the muon $g-2$ reads 
\begin{align} 
a_\mu^{U,\mathrm{w}}
\: = \: 
\big( \frac{\alpha}{\pi} 
\big)^2 \int_0^\infty \mathrm{d}t 
\:G^\mathrm{U,k} (t)\: \tilde{K} (t; m_\mu) 
\: W_I(t)  , ~~~~~~~~k=\mathrm{l},\mathrm{c},
\label{eq:UspinW}
\end{align}
where $\tilde{K}$  is the time-momentum representation kernel from~\cite{DellaMorte:2017dyu}, $G^\mathrm{U,\mathrm{l}}$ denotes local-local two-point vector correlators of non-singlet currents defined in Eq.~(\ref{eq:non-singlet}), and $G^\mathrm{U,\mathrm{c}}$ denotes conserved-local current correlator.
Both discretization choices for \uspinW 
were used to compare the statistical precision of non-perturbative QCD+QED and RM123 approaches in including IB corrections, and the mass-dependent non-singlet renormalization factor is computed by imposing that the local-local 
and conserved-local vector correlator discretizations agree at large Euclidean separations~\cite{PaolaP}.  
\begin{table}[t]
\vspace*{-1em}
   \begin{tabular}{llll}
            \hline 
            \multirow{2}{4cm}{\textbf{Observable}} & \multirow{2}{4cm}{\textbf{Kept fixed}}   & \multirow{2}{2cm}{\textbf{Physical} \\ \textbf{value}} & \multirow{2}{2cm}{\textbf{Target} \\\textbf{RC$^\star$ value}} \\& &  & \\ \hline \noalign{\vskip 1mm}
            $\sqrt{8 t_0}$& $a$& 0.415  & \halfcolor{green}{purple}{0.415 \text{fm}}  \\
            $\phi_0 = 8t_0(M_{K^\pm}^2 - M_{\pi^{\pm}}^2)$ & $m_s-m_d$   & 0.992 & \halfcolor{green}{purple}{0}  \\
            $\phi_1 = 8t_0(M_{K^{\pm}}^2 + M_{\pi^{\pm}}^2 + M_{K^{0}}^2)$ & $m_u+m_d+m_s$ & 2.26  & \halfcolor{green}{purple}{2.11}  \\
            $\phi_2 = 8t_0(M_{K^{0}}^2 - M_{K^\pm}^2)/\alpha_R$ & $m_u-m_d$  & 2.36  & {\color{green}2.36}, {\color{mag} 0} \\
            $\phi_3 = \sqrt{8t_0}(M_{D_s^\pm} + M_{D^0} + M_{D^\pm})$ & $m_c$  & 12.0  & \halfcolor{green}{purple}{12.1}  \\
            $\alpha_R$ & $e^2$ & 0.007297 & {\color{green}$\alpha^{\text{phys}}$}, {\color{mag} 0} 
        \end{tabular}
        \caption{   Hadronic renormalization scheme for \textcolor{green}{QCD+QED} and \textcolor{mag}{isoQCD}.}
        \label{tab:scheme}
\end{table}
\raggedbottom

\section{Setup for comparing the methods for including IB effects}
\label{sec:setup}

We compare the determination of \uspinW~with IB corrections including
the sea quarks effects on two ensembles generated with the~$\mathrm{openQxD}$
code~\cite{openQxDP,openQxDC}, which provides implementation of non-perturbative QCD+QED with a variety of boundary conditions. 
The QCD and QCD+QED ensembles used here, labeled \texttt{A400a00} and \texttt{A380a07}, correspond to ensembles \texttt{A400a00b324} and \texttt{A380a07b324+RW1} of Ref.~\cite{openQxDP},  and feature \cstar boundary conditions in all three spatial directions. They have the same lattice volume and the same bare QCD coupling
$\beta = 3.24,$ corresponding to the lattice spacing $a\approx 0.05\mathrm{fm}$,  
but differ in the renormalized electromagnetic coupling $\alpha_R$ and in the quark hopping parameters, $\kappa_f$. The ensemble~\texttt{A400a00}~is generated along the line of constant physics with
$\alpha_R = 0,
\kappa_u = \kappa_d = \kappa_s,$
corresponding to the SU(3)-symmetric point with $u$ and $d$ quarks degenerate.
The ensemble~\texttt{A380a07}~is generated close to the physical value of the electromagnetic coupling and features a $U$-spin symmetry between down and strange quarks.
For ensemble~\texttt{A380a07}, a nonperturbative reweighting in the bare quark mass has been applied to improve the matching to the line of constant physics. 

\begin{figure}[t]
        \vspace*{-1em} 
        \centering
           \includegraphics[width=.5\linewidth]{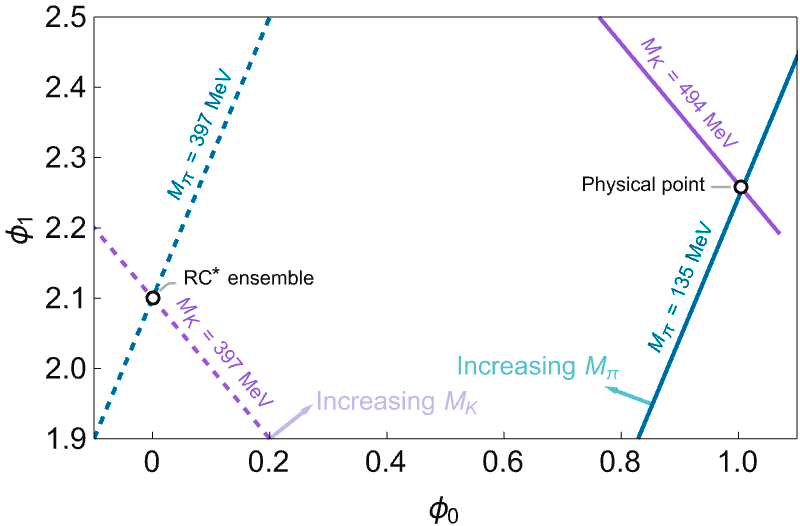}
        \caption{The "RC$^\star$ ensemble" point in parameter space, to which both ensembles used in the comparison (\texttt{A400a00} and \texttt{A380a07}) are tuned, is far from the physical point. Both points are shown here in the $\phi_0 \phi_1$-plane, where the (dashed) purple and blue lines indicate the (un)physical kaon and pion masses, respectively. 
        }
        \label{fig:physpoint}
    \end{figure}

The QCD and QCD+QED ensembles used for the comparison have unphysical pion masses, $\mathrm{M_\pi} \approx 400\:\mathrm{MeV}$. For this reason, we do not yet follow the FLAG recommendation for parametrization of isospin-symmetric QCD and IB corrections~\cite{FLAG24}, 
but rather define isospin-symmetric QCD ($\mathrm{isoQCD}$)  point by matching the quantities $\phi_i$ defined in Table~\ref{tab:scheme} to the values given in the last column of the same table, obtained from Ref.~\cite{ScaleCLS}. 
$\mathrm{isoQCD}$ 
point then corresponds to taking the limit of $\alpha \rightarrow 0, \phi_2=\mathrm{const.}$,  leading to $m_u = m_d = m_s$ in our setup (RC$^\star$ value), whose distance from the physical point in the $\phi_0 \phi_1$-plane is illustrated in Fig.~\ref{fig:physpoint}. 

    \raggedbottom

For the two ensembles, we use the Sheikholeslami-Wolhert (clover) coefficient tuned in isosymmetric QCD
$c_{SW}^{\mathrm{SU(3)}} = 2.18859,$
together with the tree-level U(1) coefficient
$c_{SW}^{\mathrm{U(1)}} = 1$
for the QCD+QED ensemble.
With these choices, $O(a)$ effects in QCD+QED are not fully removed, since this would require a non-perturbative determination of the improvement coefficients in the combined theory.
However, the goal of this work is to compare perturbative and non-perturbative implementations of QCD+QED at the same finite lattice spacing. Since the lattice action and improvement coefficients are identical in the two approaches, the comparison is consistent up to
$O(\alpha)$, 
and the specific choice of improvement coefficients is not expected to affect the conclusions of this study.

\section{Dynamical non-perturbative \gcdqedC~approach vs. \RM}
In the non-perturbative~\gcdqedC~approach,  we sample jointly $\mathrm{SU(3)}$ and $\mathrm{U(1)}$ gauge field configurations according to the probability distribution
\begin{align}
p(U,z) \propto \prod_f \mathrm{Pf} \left( CKD_f[U,z] \right)
e^{-S_{g,\mathrm{SU(3)}}[U]-S_{g,\mathrm{U(1)}}[z]},
\end{align}
where $D_f$ is the improved Wilson-Dirac operator, $U$ are as usual $\mathrm{SU(3)}$ gauge links, $z$ denote $\mathrm{U(1)}$ compact variables $z_\mu(x)=e^{iea q_{el}A_\mu(x)}$, $C$ is the charge conjugation matrix, and $K=\begin{pmatrix}
            0 & 1 \\
            1 & 0 \\
        \end{pmatrix}$. 
Here, the electromagnetic effects are incorporated directly in the generation of the gauge ensemble and the $U$-spin symmetric HVP  window observable, \uspinW, and meson masses used in defining the line of constant physics are then computed from two-point correlation functions evaluated on the dynamical QCD+QED field configurations, $O_{2\mathrm{pt}}(x,y)=\mathrm{Tr} \left\{ D_f^{-1}(y|x)\Gamma_A D_g^{-1}(x|y)\Gamma_B \right\} ,$ where 
$\Gamma_A,\Gamma_B$ are the appropriate $\Gamma$-matrices. 

\begin{table}[tb]
\centering
\vspace*{-1.2cm}
 \begin{tabular}{lll|ll}
        \\ \hline \phantom{} & \phantom{} & \phantom{} & \phantom{} & \phantom{} \\[-1.5em]
        \begin{tikzpicture}[baseline,scale=0.5]
            \coordinate (x) at (0,0);
            \draw[quark] (x) arc[start angle=-135, delta angle=90, radius=1.7] coordinate[at end] (y) node[black,below,pos=0.7]{};
            \draw (y) arc[start angle=45, delta angle=90, radius=1.7]  coordinate[pos=0.5] (p) node[black,above,pos=0.7]{};
            \draw[externalVertex] (x) circle (.1) node[left]{};
            \draw[externalVertex] (y) circle (.1) node[left]{};
            \drawTriangle{p};
            \coordinate (oi) at ($(x) - (.1,-1)$);
            \draw[vertex,transparent] (oi) circle (.1) node[right] {};
            \coordinate (of) at ($(y) + (.1,1.2)$);
            \draw[vertex,transparent] (of) circle (.1) node[right] {};
        \end{tikzpicture}
            &
        \begin{tikzpicture}[baseline,scale=0.5]
            \coordinate (x) at (0,0);
            \draw[quark] (x) arc[start angle=-135, delta angle=90, radius=1.7] coordinate[at end] (y) node[black,below,pos=0.7] {};
            \draw[quark] (y) arc[start angle=45, delta angle=90, radius=1.7]  coordinate[pos=0.3] (p) node[black,above,pos=0.7] {};
            \draw[photon] (p) arc[start angle=255, delta angle=360, radius=.34];
            \draw[externalVertex] (x) circle (.1) node[left] {};
            \draw[externalVertex] (y) circle (.1) node[right] {};
            \drawDoublePhotonInsertion{p};
            \coordinate (oi) at ($(x) - (.1,-1)$);
            \draw[vertex,transparent] (oi) circle (.1) node[right] {};
            \coordinate (of) at ($(y) + (.1,1.2)$);
            \draw[vertex,transparent] (of) circle (.1) node[right] {};
        \end{tikzpicture}
          &
        \begin{tikzpicture}[baseline,scale=0.5]
            \coordinate (x) at (0,0);
            \draw[quark] (x) arc[start angle=-135, delta angle=90, radius=1.7] coordinate[at end] (y) coordinate[pos=0.3] (p1) node[black,below,pos=0.5] {};
            \draw[quark] (y) arc[start angle=45, delta angle=90, radius=1.7] coordinate[pos=0.3] (p2) node[black,above,pos=0.5] {};
            \draw[photon] (p1) -- (p2);
            \draw[externalVertex] (x) circle (.1) node[left] {};
            \draw[externalVertex] (y) circle (.1) node[right] {};
            \drawPhotonInsertion{p1};
            \drawPhotonInsertion{p2};
            \coordinate (oi) at ($(x) - (.1,-1)$);
            \draw[vertex,transparent] (oi) circle (.1) node[right] {};
            \coordinate (x) at (4,0);
            \draw[quark] (x) arc[start angle=-135, delta angle=90, radius=1.7] coordinate[at end] (y) node[black,below,pos=0.9] {};
            \draw[quark] (y) arc[start angle=45, delta angle=90, radius=1.7] coordinate[pos=0.3] (p1) coordinate[pos=0.7] (p2) node[black,above,pos=0.9] {};
            \draw[photon] (p1) .. controls ($(p1)!0.2!(p2) - (0,.5)$) and ($(p1)!0.8!(p2) - (0,.5)$) .. (p2);
            \draw[externalVertex] (x) circle (.1) node[left] {};
            \draw[externalVertex] (y) circle (.1) node[right] {};
            \drawPhotonInsertion{p1};
            \drawPhotonInsertion{p2};
            \coordinate (of) at ($(y) + (.1,1.2)$);
            \draw[vertex,transparent] (of) circle (.1) node[right] {};
        \end{tikzpicture}
          &
        \begin{tikzpicture}[baseline,scale=0.5]
            \coordinate (x) at (0,0);
            \draw[quark] (x) arc[start angle=-135, delta angle=90, radius=1.7] coordinate[at end] (y) node[black,below,pos=0.7] {};
            \draw[quark] (y) arc[start angle=45, delta angle=90, radius=1.7]  coordinate[pos=0.3] (p) node[black,above,pos=0.7] {};
            \draw[photon] (p) .. controls ($(p)!0.2!(x) - (0,.5)$) and ($(p)!0.8!(x)$) .. (x);
            \draw[externalVertex] (y) circle (.1) node[right] {};
            \drawPhotonInsertion{p};
            \drawVertexPhotonInsertion{x};
            \coordinate (oi) at ($(x) - (.1,-1)$);
            \draw[vertex,transparent] (oi) circle (.1) node[right] {};
            \coordinate (of) at ($(y) + (.1,1.2)$);
            \draw[vertex,transparent] (of) circle (.1) node[right] {};
        \end{tikzpicture}
          &
        \begin{tikzpicture}[baseline,scale=0.5]
            \coordinate (x) at (0,0);
            \draw[quark] (x) arc[start angle=-135, delta angle=90, radius=1.7] coordinate[at end] (y) node[black,below,pos=0.7] {};
            \draw[quark] (y) arc[start angle=45, delta angle=90, radius=1.7];
            \draw[externalVertex] (y) circle (.1) node[right] {};
            \draw[photon] (x) arc[start angle=0, delta angle=360, radius=.34];
            \drawDoubleVertexPhotonInsertion{x};
            \coordinate (oi) at ($(x) - (.1,-1)$);
            \draw[vertex,transparent] (oi) circle (.1) node[right] {};
            \coordinate (of) at ($(y) + (.1,1.2)$);
            \draw[vertex,transparent] (of) circle (.1) node[right] {};
        \end{tikzpicture}
        \\ \hline \phantom{} & \phantom{} & \phantom{} & \phantom{} & \phantom{} \\[-1.5em]
        \phantom{} &
        \phantom{} &
        \begin{tikzpicture}[baseline,scale=0.5]
            \coordinate (x) at (2,0);
            \draw[quark] (x) arc[start angle=-135, delta angle=90, radius=1.7] coordinate[at end] (y) node[black,below,pos=0.7] {};
            \draw[quark] (y) arc[start angle=45, delta angle=90, radius=1.7]  coordinate[pos=0.3] (p1);
            \coordinate (z) at ($(x)!0.5!(y) + (0,1)$);
            \draw[quark] (z) arc[start angle=0, delta angle=360, radius=.34];
            \draw[photon] (p1) .. controls ($(p1) + (0,.25)$) and ($(z) + (.25,0)$) .. (z);
            \drawPhotonInsertion{p1};
            \drawPhotonInsertion{z};
            \draw[externalVertex] (x) circle (.1) node[right] {};
            \draw[externalVertex] (y) circle (.1) node[right] {};
            \coordinate (oi) at ($(x) - (2.1,-1)$);
            \draw[vertex,transparent] (oi) circle (.1) node[right] {};
            \coordinate (of) at ($(y) + (2.1,1.5)$);
            \draw[vertex,transparent] (of) circle (.1) node[right] {};
        \end{tikzpicture}
          &
        \begin{tikzpicture}[baseline,scale=0.5]
            \coordinate (x) at (0,0);
            \draw[quark] (x) arc[start angle=-135, delta angle=90, radius=1.7] coordinate[at end] (y) node[black,below,pos=0.7] {};
            \draw[quark] (y) arc[start angle=45, delta angle=90, radius=1.7];
            \coordinate (z) at ($(x)!0.5!(y) + (0,1)$);
            \draw[quark] (z) arc[start angle=180, delta angle=360, radius=.34];
            \draw[photon] (x) .. controls ($(x) + (0,.5)$) and ($(z) - (.25,0)$) .. (z);
            \drawVertexPhotonInsertion{x};
            \drawPhotonInsertion{z};
            \draw[externalVertex] (y) circle (.1) node[right] {};
            \coordinate (oi) at ($(x) - (.1,-1)$);
            \draw[vertex,transparent] (oi) circle (.1) node[right] {};
            \coordinate (of) at ($(y) + (.1,1.5)$);
            \draw[vertex,transparent] (of) circle (.1) node[right] {};
        \end{tikzpicture}
          &
        \phantom{}
        \\ \phantom{} & \phantom{} &\phantom{} & \phantom{} & \phantom{} \\[-1.4em]
        \begin{tikzpicture}[baseline,scale=0.5]
            \coordinate (x) at (0,0);
            \draw[quark] (x) arc[start angle=-135, delta angle=90, radius=1.7] coordinate[at end] (y) node[black,below,pos=0.7] {};
            \draw[quark] (y) arc[start angle=45, delta angle=90, radius=1.7];
            \coordinate (z) at ($(x)!0.5!(y) + (0,.8)$);
            \draw[quark] (z) arc[start angle=270, delta angle=360, radius=.34];
            \drawTriangle{z};
            \draw[externalVertex] (x) circle (.1) node[right] {};
            \draw[externalVertex] (y) circle (.1) node[right] {};
            \coordinate (oi) at ($(x) - (.3,-1)$);
            \draw[vertex,transparent] (oi) circle (.1) node[right] {};
            \coordinate (of) at ($(y) + (.1,1.5)$);
            \draw[vertex,transparent] (of) circle (.1) node[right] {};
        \end{tikzpicture}
            &
        \begin{tikzpicture}[baseline,scale=0.5]
            \coordinate (x) at (0,0);
            \draw[quark] (x) arc[start angle=-135, delta angle=90, radius=1.7] coordinate[at end] (y) node[black,below,pos=0.7] {};
            \draw[quark] (y) arc[start angle=45, delta angle=90, radius=1.7];
            \coordinate (z) at ($(x)!0.5!(y) + (0,.8)$);
            \draw[quark] (z) arc[start angle=270, delta angle=360, radius=.34] coordinate[pos=0.25] (z1) coordinate[pos=0.75] (z2);
            \draw[photon] (z2) arc[start angle=0, delta angle=360, radius=.34];
            \draw[externalVertex] (x) circle (.1) node[right] {};
            \draw[externalVertex] (y) circle (.1) node[right] {};
            \drawDoublePhotonInsertion{z2};
            \coordinate (oi) at ($(x) - (.3,-1)$);
            \draw[vertex,transparent] (oi) circle (.1) node[right] {};
            \coordinate (of) at ($(y) + (0.1,1.5)$);
            \draw[vertex,transparent] (of) circle (.1) node[right] {};
        \end{tikzpicture}
          &
        \begin{tikzpicture}[baseline,scale=0.5]
            \coordinate (x) at (0,0);
            \draw[quark] (x) arc[start angle=-135, delta angle=90, radius=1.7] coordinate[at end] (y) node[black,below,pos=0.7] {};
            \draw[quark] (y) arc[start angle=45, delta angle=90, radius=1.7];
            \coordinate (z) at ($(x)!0.5!(y) + (0,.8)$);
            \draw[quark] (z) arc[start angle=270, delta angle=360, radius=.34] coordinate[pos=0.25] (z1) coordinate[pos=0.75] (z2);
            \draw[photon] (z1) -- (z2);
            \draw[externalVertex] (x) circle (.1) node[right] {};
            \draw[externalVertex] (y) circle (.1) node[right] {};
            \drawPhotonInsertion{z1};
            \drawPhotonInsertion{z2};
            \coordinate (oi) at ($(x) - (.1,-1)$);
            \draw[vertex,transparent] (oi) circle (.1) node[right] {};
            \coordinate (x) at (4,0);
            \draw[quark] (x) arc[start angle=-135, delta angle=90, radius=1.7] coordinate[at end] (y) node[black,below,pos=0.7] {};
            \draw[quark] (y) arc[start angle=45, delta angle=90, radius=1.7];
            \coordinate (z1) at ($(x)!0.8!(y) + (0,0.8)$);
            \coordinate (z2) at ($(x)!0.2!(y) + (0,0.8)$);
            \draw[quark] (z1) arc[start angle=270, delta angle=360, radius=.34] coordinate[pos=0.75] (p1);
            \draw[quark] (z2) arc[start angle=270, delta angle=360, radius=.34] coordinate[pos=0.25] (p2);
            \draw[photon] (p1) -- (p2);
            \draw[externalVertex] (x) circle (.1) node[right] {};
            \draw[externalVertex] (y) circle (.1) node[right] {};
            \drawPhotonInsertion{p1};
            \drawPhotonInsertion{p2};
            \coordinate (of) at ($(y) + (.1,1.5)$);
            \draw[vertex,transparent] (of) circle (.1) node[right] {};
        \end{tikzpicture}
          &
        \phantom{} &
        \phantom{} 
        \\ \hline
    \end{tabular}
\caption{Feynman diagrams contributing to the IB corrections to \uspin~in the \RM~approach, grouped into valence–valence diagrams (first row), sea–valence diagrams (second row), and the sea–sea diagrams (third row).
The right column shows the additional diagrams that arise when the point-split current is inserted at the sink. Only valence-quark connected diagrams contributing to the $U$-spin symmetric observable considered in this work are shown. For more details on the diagrams structure and evaluation, see~\cite{PaolaP} and~\cite{PaolaThesis}.
}
\label{tab:Table2}
\end{table}

In the \RM~strategy, we sample field configurations from QCD probability distribution, and incorporate IB corrections through an expansion of the quark propagator and other operators.
QED and strong IB effects are obtained by perturbatively expanding around the isoQCD point defined in Sec.~\ref{sec:setup}. Here, the expansion of the probability density can be written as 
\begin{align}
p(U,A) \propto \prod_f \mathrm{Pf} \left( CKD_f^{(0)}[U]\right)
e^{-S_{g,\mathrm{SU(3)}}[U]} e^{-S_\gamma[A]}
\left[1+O(\Delta D_f[U,A])\right],
\end{align}
where $S_{\gamma}[A]$ denotes a non-compact formulation of the $\mathrm{U(1)}$ gauge action $S_\gamma [A] =\frac{a^4}{4} \sum_x \sum_{\mu,\nu} F_{\mu\nu}^2(x)$, $F_{\mu\nu}=\partial_\mu A_\nu-\partial_\nu A_\mu$, ~$\mu,\nu=0,\dots3$. 
The leading order correlator is computed as
$$  O^{(0)}_{2\mathrm{pt}}(x,y) =
\mathrm{Tr}
\left\{
\left({D_f^{(0)}}\right)^{-1}(y|x)\Gamma_A
\left({D_g^{(0)}}\right)^{-1}(x|y)\Gamma_B
\right\},
$$
while the IB corrections are obtained as proposed in~\cite{RM123} from the expansion of the propagators or the $\Gamma$ matrices 
\begin{equation*}
\begin{aligned}
D_f^{-1}: &\;
\begin{tikzpicture}[baseline=-.1cm,scale=0.6]
\draw[black,quark,double] (-0.7,0) -- (0.7,0) node[black,below,pos=0.5] {$f$};
\end{tikzpicture}
= 
\begin{tikzpicture}[baseline=-.1cm,scale=0.6]
\draw[black,quark] (-0.7,0) -- (0.7,0) node[black,below,pos=0.5] {$f$};
\end{tikzpicture}
+ \Delta m_f
\begin{tikzpicture}[baseline=-.1cm,scale=0.6]
    \draw[black,quark] (-0.7,0) -- (0,0) node[black,below,pos=0.5] {$f$};
    \draw[black,quark] (0,0) -- (0.7,0) node[black,below,pos=0.5] {$f$};
    \draw[fill=red, draw=black] (0,0.15) -- (-0.125,-0.075) -- (0.125,-0.075) -- cycle;
\end{tikzpicture}
\, + e q_f \, 
\begin{tikzpicture}[baseline=-.1cm,scale=0.6]
    \draw[black,quark] (-0.7,0) -- (0,0) node[black,below,pos=0.5] {$f$};
    \draw[black,quark] (0,0) -- (0.7,0) node[black,below,pos=0.5] {$f$};
    \draw[black,photon] (0,0) -- (0,0.7);
    \draw[fill=green, draw=black] (-0.1,-0.1) rectangle (0.1,0.1);
\end{tikzpicture}
\, + \frac{1}{2}e^2 q_f^2 \,
\begin{tikzpicture}[baseline=-.1cm,scale=0.6]
    \draw[black,quark] (-0.7,0) -- (0,0) node[black,below,pos=0.5] {$f$};
    \draw[black,quark] (0,0) -- (0.7,0) node[black,below,pos=0.5] {$f$};
    \draw[black,photon] (0,0) -- (-.5,0.7);
    \draw[black,photon] (0,0) -- (0.5,0.7);
    \draw[fill=blue!60!cyan, draw=black, rotate=45] (-0.1,-0.1) rectangle (0.1,0.1);
\end{tikzpicture}, 
\\
\Gamma_{\tilde{V}}:& \;
\begin{tikzpicture}[baseline=-.1cm,scale=0.6]
        \draw[black,quark] (0,0) -- (.7,0.7) node[above,pos=0.7] {};
        \draw[black,quark] (.7,-0.7) -- (0,0) node[below,pos=0.7] {};
        \draw[vertex] (0,0) circle (.1) node[left] {};
\end{tikzpicture},
    \, = \,
        \begin{tikzpicture}[baseline=-.1cm,scale=0.6]
        \draw[black,quark] (0,0) -- (.7,0.7) node[above,pos=0.7] {};
        \draw[black,quark] (.7,-0.7) -- (0,0) node[below,pos=0.7] {};
        \fill[black] (0,0) circle (.1);
    \end{tikzpicture}
    \, + \,
    e q_f \,
\begin{tikzpicture}[baseline=-.1cm,scale=0.6]
        \draw[black,quark] (0,0) -- (.7,0.7) node[above,pos=0.5] {};
        \draw[black,quark] (.7,-0.7) -- (0,0) node[below,pos=0.5] {};
        \draw[photon] (-.7,0) -- (0,0);
        \node[regular polygon, regular polygon sides=5, draw=black, fill=orange, minimum size=5pt, inner sep=0pt, outer sep=0pt] at (0,0) {};
\end{tikzpicture}
\, + \frac{1}{2}e^2 q_f^2 \,
\begin{tikzpicture}[baseline=-.1cm,scale=0.6]
        \draw[black,quark] (0,0) -- (.7,0.7) node[above,pos=0.5] {};
        \draw[black,quark] (.7,-0.7) -- (0,0) node[below,pos=0.5] {};
        \draw[photon] (-.7,0.7) -- (0,0);
        \draw[photon] (-.7,-0.7) -- (0,0);
        \node[star, star points=5, draw=black, fill=yellow, minimum size=5pt, inner sep=0pt, outer sep=0pt] at (0,0) {};
\end{tikzpicture}.
\end{aligned}
\end{equation*}
The Feynman diagrams that need to be computed, in order to include all IB corrections to the valence-quark connected two-point function relevant for \uspinW up to next-to-leading order in the expansion parameters, are listed in Table~\ref{tab:Table2}.
\raggedbottom

\section{Strategies for comparing the two methods to include IB corrections}
\label{sec:strategies}
    \begin{figure}
    \vspace*{-2em}
        \centering
        \includegraphics[width=0.7\linewidth]{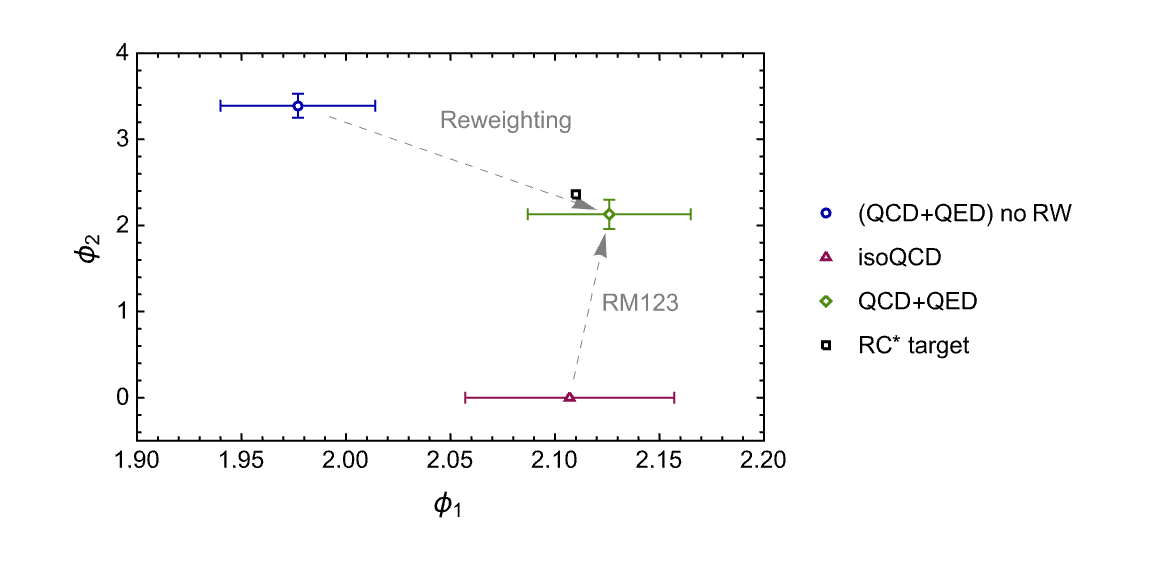}
        \vspace*{-1emd}
        \caption{Mass reweighting was applied to the non-perturbative QCD+QED ensemble \texttt{A380a07} to match the line of constant physics. We found negligible impact on the statistical uncertainties of the observables.}
        \label{fig:matching}
    \end{figure}
    \raggedbottom
To compare the non-perturbative QCD+QED and the RM123 determinations of IB effects,  we apply mass reweighting (RW) to match between isoQCD and QCD+QED ensembles, as illustrated in Fig.~\ref{fig:matching}. To this end, we reweight the QCD+QED ensemble to improve its agreement with the target RC$^\star$ value defined in Table.~\ref{tab:scheme}. The reweighting factors computed in earlier work~\cite{openQxDP2} were shown not to increase the statistical uncertainties. The matching conditions are expressed in terms of the dimensionless quantities $\phi_i$
defined in Table~\ref{tab:scheme}, and gradient flow scale $t_0$ is set to the central value of the  $N_f=3$ CLS determination~\cite{ScaleCLS}.  
The reweighting factor is defined as $\mathcal{R}=\frac{\prod_f\mathrm{Pf} \left( CKD_f \right)}{\prod_f\mathrm{Pf} \left( CKD_f^{(0)} \right)}$. 

We consider two strategies to perform a comparison of the non-perturbative QCD+QED calculation with the \RM~approach.
In the first strategy, the comparison is carried out at fixed bare parameters. In this case, in the perturbative expansion we use the exact mass shifts obtained from the difference of hopping parameters 
\begin{align}
a\Delta m_{\mathrm u} = -0.00476435, \quad
a\Delta m_{\mathrm {d,s}} = -0.00077259,\quad
a\Delta m_{\mathrm c} = -0.00682735,
\label{eq:exactshifts}
\end{align}
and observables 
are compared according to
\begin{align}
X^\mathrm{nonpert.} ~~~~ \text{vs.} ~~\quad
X(0)
+e^2 \partial_{e^2}X(0)
+\sum_{f=\mathrm{u,d,s,c}} \Delta m_f \partial_{m_f}X(0),
\end{align}
for all quantities entering the analysis, namely $t_0$, $\phi_i$, and $a^{\mathrm{U,w}}_{\mu}$.

The second strategy for comparison follows fixed line of constant physics. In this case, the quark mass shifts in the \RM~approach are determined by solving the system of equations corresponding to the tuning conditions
\begin{align}
\phi^{(0)}_i + e^2\partial_{e^2} \phi_i
+ \sum_f \Delta m_f , \partial_{m_f} \phi_i
- \Delta_L\phi_i
  = \phi_i^{\star},
  \label{eq:conditions}
\end{align}
  such that the physical point is matched after including finite volume corrections. This way, the uncertainties on the tuned mass shifts propagate to the prediction of $a^{\mathrm{U,w}}_{\mu}$, while in the non-perturbative dynamical QCD+QED calculation the tuning error is propagated as
\begin{align}
(\mathrm{d}{a^{\mathrm{U,w}}_\mu})^2
= \sum_f \left(\partial_{m_f} a_{\mu}^{\mathrm{U,w}} \right)^2 
(\mathrm{d}m_f)^2,
\qquad \mathrm{d}m_f(\vec{\phi}) = \sum_i (J^{-1})_{fi},\mathrm{d}\phi_i,
\end{align}
where the inverse Jacobian, $J^{-1}$, tracks change of variables from the hadronic quantities to the bare quark masses. In the second strategy, we compare directly the final observable $a^{\mathrm{U,w}}_{\mu}$, and the derivatives $\mathrm{d}\phi_i$ and $\mathrm{d}{a^{\mathrm{U,w}}_\mu}$
are computed at the isoQCD point, which is an
approximation valid at first order in the bare parameters expansion.

\section{Comparing the dynamical non-perturbative QCD+QED and the \RM~calculations}
In the non-perturbative dynamical \gcdqedC~setup, the computation of the \uspinW including the IB requires computing a single Wick contraction, and it agrees with the computation of the same observable on the isoQCD ensemble, $\texttt{A400a00}$,  without considering the IB corrections
\begin{align}
a_\mu^\mathrm{U,w}\times 10^{11} &=
\begin{cases}
 \textrm{local-local} \quad   \textrm{conserved-local} & ~\\
1083(6)  \quad \qquad 1086(5)  & \quad\textrm{isoQCD},\\
  1082(7) \quad \qquad 1085(7)  & \quad\textrm{non-perturbative {QCD+QED}.}
\end{cases}
\label{eq:isoQCD}
\end{align}

\begin{table}[t]
\vspace*{-1em}
\centering
\begin{tabular}{ c c c}
        \toprule
        Class & $\ell=\mathrm l$  & $\mathrm c$\\
        \midrule 
             \textrm{valence-valence} & -6.6(4)  & -6.6(4)  \\
             \textrm{sea-valence} &  -0.24(1)&  -0.24(1)\\
             \textcolor{mag}{\textrm{sea-sea}} & \textcolor{mag}{14(18)} & \textcolor{mag}{13(18)}  \\
        \midrule 
        $\delta a_{\mu}^{\mathrm{U,w}}$ & 7(18) & 6(18) \\
        \bottomrule
    \end{tabular}
    \caption{Decomposition of the IB corrections to \uspinW onto valence-valence, sea-valence, and sea-sea corrections classified in Table~\ref{tab:Table2}. The uncertainty in the \RM~approach is dominated by the sea-sea effects.}
\label{tab:decomposition}
\end{table}
\raggedbottom

In the first of the two strategies for comparing non-perturbative dynamical QCD+QED with the~\RM~setup, the uncertainties related to the tuning to the lines of constant physics
can be computed separately from the total uncertainty in the IB correction to \uspinW.  To this end, we can compute the correction to the scale by expanding the condition
$ t_0^2 \langle E(t_0)\rangle = 0.3 $
around $t_0^{(0)}$, yielding 
\begin{align} 
\delta t_0 =
\frac{-t_0 \Delta\varepsilon \cdot \nabla \langle E(t_0)\rangle}
{2\langle E(t_0)\rangle
+ t_0 \langle d/dt E(t)|_{t=t_0}\rangle},~~~~~~~
  \Delta\varepsilon =
  (e^2,\Delta m_\mathrm{u},\Delta m_\mathrm{d},\Delta m_\mathrm{s},\Delta m_\mathrm{c}),
\end{align}
 where only sea-quark effects contribute to $\nabla \langle E(t_0)\rangle$.
  The resulting derivatives in lattice units are
\begin{align} 
  \hat t_0^{(0)} = 7.400(69), \quad
  \partial_{am_{\mathrm u}} \hat t_0 = -76(24), \quad
  \partial_{am_{\mathrm c}} \hat t_0 = -26.5(8.1), \quad
  \partial_{e^2}\hat t_0 = -6.1(1.9),
\end{align}
  leading to the scale correction
\begin{align} 
  \delta \hat t_0 = 0.10(3),
  \qquad
  \frac{\delta a}{a^{(0)}}
  =
- \frac{\delta t_0}{2t_0^{(0)}}
  =
  -0.0069(19).
\end{align}
Similarly, the corrections to the quantities $\phi_i$ that define the lines of constant physics are given by
\begin{align} 
\delta \phi_i= e^2 \partial_{e^2}\phi_i + \sum_f \Delta m_f \partial_{m_f}\phi_i -\Delta_L \phi_i,
\end{align}
where $\Delta_L \phi_i$ are obtained from the universal finite volume corrections to the meson masses in \cstar setup in three spatial directions~\cite{Lucini:2015hfa}
\begin{align} 
\Delta M(L) = \frac{e^2 q^2}{4\pi}
\left(
\frac{\zeta(1)}{2L} +
\frac{\zeta(2)}{\pi M L^2}
\right), 
\qquad \zeta(1)=-1.7475645946, \quad \zeta(2)=-2.5193561521.
\end{align}
Using the exact input values for $\Delta m_f$ given in Eq.~(\ref{eq:exactshifts}), the resulting corrections to $\phi_i$ at fixed bare parameters are
\begin{align} 
\delta \phi_0=0.0,
\qquad
\delta \phi_1=0.084(82),
\qquad
\delta \phi_2=2.52(14),
\qquad
\delta \phi_3=0.053(47),
\end{align}
were $\phi_0=0$ is enforced by $U$-spin symmetry.

The correction to the HVP-related observable of interest
$a_\mu^{\mathrm{U,w}}$
can be decomposed into correlator  ($G$), renormalization ($Z$), and scale ($a$) corrections, 
$
\delta a_\mu^{\mathrm{U,w}}
=
\delta_G a_\mu^{\mathrm{U,w}}
+
\delta_Z a_\mu^{\mathrm{U,w}}
+
\delta_a a_\mu^{\mathrm{U,w}}.
$
The individual terms, in units of $10^{-11}$, for the specified bare mass shifts amount to
\begin{align*}
&\delta_X a_\mu^{\mathrm{U,w}} \times 10^{11}  & \textrm{local-local}  &~&\textrm{conserved-local} \\
~&\delta_G a_\mu^{\mathrm{U,w}}  \qquad & 11(16)   &\qquad &  \qquad  \qquad \qquad  ~12(17), \\
~&\delta_Z a_\mu^{\mathrm{U,w}}  \qquad & ~~~~3.8(3.4)  &\qquad &  1.9(1.7), \\
~&\delta_a a_\mu^{\mathrm{U,w}}  \qquad &  -8.4(2.3)  &\qquad &   ~~-8.4(2.4).
\end{align*}
leading to the total \uspinW correction estimates for the two discretization of the $U$-spin current
\begin{align*}
\delta a_\mu^\mathrm{U,w}\times 10^{11} &=
\begin{cases}
7(18)   & \quad\textrm{local-local},\\
6(18)  & \quad\textrm{conserved-local.}
\end{cases}
\end{align*}
The results for isoQCD+RM123 at fixed bare parameters are then obtained by adding the computed IB corrections to the isoQCD result reported in Eq.~(\ref{eq:isoQCD}).
A decomposition of the correction to
\uspinW~given in Table~\ref{tab:decomposition} shows that the dominant source of uncertainty originates from sea–sea contributions, while valence–valence and sea–valence terms have a very good signal using the same statistics.

\begin{table}[t]
\vspace*{-1em}
\centering
\small
{\begin{tabular}{ccccc}
    \toprule
     &  \multicolumn{2}{c}{Fixed bare parameters} & \multicolumn{2}{c}{Line of const. physics}\\
      & $ \textrm{local-local}$ & $ \textrm{conserved-local}$ & $ \textrm{local-local}$  & $ \textrm{conserved-local}$ \\
     \midrule 
     isoQCD & 1083(6) & 1086(5) &  1083(6) & 1086(5) \\
     isoQCD+RM123$|_{\mathrm{eq}}$ & 1078(5) & 1080(5) & 1084(5) & 1087(5)  \\
     \midrule
        isoQCD+RM123   &  1090(18) & 1092(18)  & 1093(20) & 1094(21)              \\
        non-perturbative QCD+QED  & 1082(7) & 1085(7) & 1082(8) & 1085(7)\\
        \bottomrule
\end{tabular}}
\caption{Comparison of the dynamical non-perturbative QCD+QED and \RM-computed IB-corrected \uspinW following the two strategies described in Sec.~\ref{sec:strategies}.}
\label{tab:final}
\end{table}
\raggedbottom

The result for the total \uspinW is confirmed in the computation following the second comparison strategy. Here, the quark mass shifts and their uncertainties determined by fixing the line of constant physics as in Eq.~(\ref{eq:conditions}) read
\begin{align}
a\Delta m_{\mathrm u} = -0.00477(17), \quad
a\Delta m_{\mathrm {d,s}} =  -0.00082(17),\quad
a\Delta m_{\mathrm c} =  -0.0083(28),
\label{eq:LCPshifts}
\end{align}
and give IB-corrected value for both discretizations of the current 
which is in agreement with the initial comparison strategy based on fixed bare parameters. The comparison of the two strategies, showing first the isoQCD alongside electroquenched result, augmented with the sea-quark effects (isoQCD+RM123) and a separate dynamical non-perturbative QCD+QED result, is shown in Table~\ref{tab:final}. 

\section{Summary and outlook}
We reported on the first systematic comparison between dynamical non-perturbative QCD+QED simulations with \cstar boundary conditions and the perturbative RM123 expansion applied to the two ensembles with the same 
fermion discretization, lattice volume, and similar bare parameters.
We performed calculations at a single lattice spacing, and unphysical pion mass, focusing on the intermediate window contribution for a flavour non-singlet current to the HVP,
$a^{\mathrm{U,w}}_{\mu}$, with an aim to compare total uncertainty in both
methods with a fixed number of gauge field configurations. Preliminary results of this comparison, without including sea-quark effects, were reported in~\cite{LetiziaProc}, and further details on the updated study presented here can be found in~\cite{PaolaP,PaolaThesis}.

In the $U$-spin symmetric setup, chosen to eliminate valence quark-line disconnected diagrams, the two methods for including IB corrections give consistent results for the intermediate window HVP.  The fully non-perturbative, dynamical QCD+QED, calculation achieves consistently $2.5-3$  times smaller statistical uncertainties,
for both discretizations and for both comparison strategies, while in the \RM~approach the statistical uncertainty is completely dominated by the sea-sea contributions. 

The presented results provide a controlled benchmark for future calculations including sea-quark effects  within the \rcstar collaboration programme. Future work will extend this comparison to larger volumes and to ensembles closer to the physical pion mass, where improved variance-reduction techniques~\cite{Gruber:2024cos,TimProceedings} are expected to be essential. Additional lattice spacings will be required to perform continuum extrapolations, and the study will be generalized to the full HVP, including a study of valence quark-line disconnected contributions and finite-volume effects with \cstar boundary conditions. Moving away from the $U$-spin symmetric point will require the inclusion of valence quark-line disconnected diagrams also in the RM123 framework, for which dedicated noise-reduction methods are currently under development.

\acknowledgments
We are grateful to the members of the RC$^\star$ collaboration and the Muon g-2 Theory IWnitiative for insightful discussions. 
We acknowledge access to the Swiss National Supercomputing Centre (CSCS) (eth8, c21, and s1196) and to NHR@ZIB and NHR@G\"ottingen as part of the NHR infrastructure (projects bep00085, bep00102 and bep00116). This research was supported by  SNSF via Project No. 200021\_200866, the Platform for Advanced Scientific Computing (PASC), and in part by grant NSF PHY-2309135 to the Kavli Institute for Theoretical Physics. JK, MKM and ICP acknowledge the support from the Horizon Europe project interTwin, funded by the European Union Grant Agreement Number https://cordis.europa.eu/project/id/101058386. F.M. and N.T. are supported by the Italian Ministry of University and Research and the European Union (EU)--Next Generation EU, Mission 4, Component 1, PRIN 2022, CUP F53D23001480006 and CUP D53D23002830006. A.C.’s research is funded by the Deutsche Forschungsgemeinschaft (DFG, German Research Foundation)--Projektnummer 417533893/GRK2575 “Rethinking Quantum Field Theory”.

\bibliographystyle{JHEP}
\bibliography{references}
\end{document}